\documentstyle[preprint,aps]{revtex}

\begin{document}
\setlength{\topmargin}{0in}
\setlength{\textheight}{8.5in}
\draft
\preprint{}
\title{Coupled Hartree-Fock-Bogoliubov kinetic 
equations \\  for a trapped Bose gas}
\author{Milena Imamovi\'{c}-Tomasovi\'{c} and Allan Griffin}
\address{Department of Physics, University of Toronto 
\\ Toronto, Ontario, Canada 
M5S 1A7}
\date{\today}
\maketitle
\begin{abstract}
Using the Kadanoff-Baym non-equilibrium Green's function formalism, we derive the 
self-consistent Hartree-Fock-Bogoliubov (HFB) collisionless kinetic equations 
and the associated equation of motion for the condensate wavefunction for
a trapped Bose-condensed 
gas. Our work generalizes earlier work by Kane and Kadanoff (KK) 
for a uniform Bose gas. We include the off-diagonal (anomalous) pair correlations, and 
thus we have to introduce an  off-diagonal distribution 
function in addition to the normal (diagonal) distribution function.
 This results in two coupled kinetic equations.  
If the off-diagonal distribution function can be neglected 
as a higher-order contribution, we obtain the semi-classical kinetic equation 
recently used by  Zaremba, Griffin and  Nikuni (based on the simpler 
Popov approximation). We discuss  the  
static local equilibrium solution of our coupled HFB kinetic equations 
within the semi-classical approximation. We also 
verify that a solution is the  rigid in-phase oscillation of the equilibrium condensate and 
non-condensate density profiles, oscillating  with the trap frequency.   
\end{abstract}

\newcommand{\r}{{\bf r}}
\newcommand{\up}{\stackrel{<}{>}}
\newcommand{\ba}{\begin{eqnarray}}
\newcommand{\ea}{\end{eqnarray}}
\newcommand{\be}{\begin{equation}}
\newcommand{\ee}{\end{equation}}

\section{Introduction}
In their classic book, Kadanoff and Baym (KB) \cite{kb} developed a 
systematic way of deriving quantum Boltzmann equations using 
non-equilibrium Green's functions. KB derived generalized Boltzmann 
equations for 
interacting Bose and Fermi gases in the normal phase (no broken 
symmetry). The advantage of this approach is that one can generate 
kinetic equations starting from a well-defined single-particle 
self-energies using functional derivatives. The KB method was 
generalized by Kane and Kadanoff (KK) \cite{kk} to deal with 
a Bose-condensed gas, with the goal of using the 
resulting kinetic equations to derive the two-fluid hydrodynamic 
equations of Landau. The KK analysis worked with diagonal and 
off-diagonal self-energies, and included terms of second order in 
the two-particle interaction (the Beliaev second order
self-energy diagrams \cite{shi}). This gives kinetic equations which include
the effect of collisions between the atoms.

In the present paper, we use the KK approach to deal with a trapped 
Bose-condensed gas. For simplicity, in the present paper we limit 
ourselves to Hartree-Fock (HF) self-energies and thus the kinetic 
equations we obtain are in the collisionless 
approximation\cite{kb}. 
Because we include both normal and 
anomalous pair correlations, our self-energies give the so-called 
Hartree-Fock-Bogoliubov (HFB) approximation, as 
reviewed by Griffin \cite{hfb}. While the HFB approximation in a 
uniform gas is known to lead to a single-particle spectrum with an
energy gap in the long wavelength limit, it is a ``conserving'' approximation 
which will generate two-particle response functions which satisfy 
conservation laws (see Refs. \cite{kb,kbphyrev} and Section VI for 
further discussion).

Our main formal result is given in Section III, namely two coupled kinetic
 equations for the  diagonal ($f_{1}$)
and off-diagonal ($f_{2}$) distribution functions for the excited atoms, in addition to 
an equation of motion for the condensate order parameter. If $f_{2}$ can be ignored, 
we obtain the semi-classical collisionless kinetic equation for 
$f_{1}$ which has 
been used in recent discussions of the non-equilibrium properties 
of the non-condensed atoms \cite{zgn,stoof}. In order to gain some 
insight into our  HFB kinetic equations, in Section IV we  
discuss  the static local equilibrium solutions within the semi-classical 
approximation and verify that these satisfy our  
equations. In section V, we show that 
our equations have a solution which corresponds to a  a rigid oscillation of the equilibrium 
density profiles (normal and anomalous), with a frequency equal to the 
parabolic trap frequency. This generalized Kohn mode is generic but 
how it arises as a solution of our HFB kinetic equations gives one insight into 
their structure.
 
\section{Derivation of HFB equations}
In terms of Bose quantum field operators, the many-body Hamiltonian 
($\hat{K}=\hat{H}-\mu_{0}\hat{N}$) describing
interacting Bosons confined 
by an external potential $U_{ext}({\bf r})$ is given by:
\begin{equation}
\hat{K}=\int d{\bf r} {\psi}^{\dag}({\bf 
r})\left[-\frac{1}{2m}\nabla^{2}_{\r}+U_{ext}({\bf r})-\mu_{0}\right] 
{\psi}({\bf r}) +\frac{1}{2} \int d{\bf r} d{\bf r'}
 {\psi}^{\dag}({\bf r})  {\psi}^{\dag}({\bf r'}) 
v( {\bf r} - {\bf r'}) {\psi}({\bf r}) {\psi}({\bf r'}).
\end{equation}
For a discussion of the properties of a dilute Bose gas at low 
temperatures the two-body interaction $v({\bf r}-{\bf r'})$ can be 
effectively treated using the s-wave approximation: 
$v({\bf r})=g\delta({\bf r})$, where $g=4\pi a/m$ (we set $\hbar=1$ throughout 
this article).
We shall be interested in the $2\times 2$ matrix single-particle 
Green's function defined by \cite{kk,fw}
\begin{equation}
\hat{g}(1,1';U)=-i\left ( \begin{array}{cc}\langle T 
 \psi(1) \psi^{\dag}(1')\rangle \hspace*{5mm}\langle T \psi(1) 
 \psi(1')\rangle\\ 
\langle T \psi^{\dag}(1) \psi^{\dag}(1')\rangle
\hspace*{5mm} \langle T \psi^{\dag}(1) \psi(1')\rangle \end{array} \right ).
\end{equation}
Here, {\it T} represents the time-ordering 
operation and we use the usual abbreviated notation \cite{kb}, 1$\equiv ({\bf {r}} 
t)$ and $1'\equiv ({\bf r'} t'$).  
We separate out the condensate part of the field operator in the usual fashion 
\begin{equation}
 \psi({\bf {r}})=\langle\psi({\bf r})\rangle_{t}+\tilde{\psi}({\bf r}),
\end{equation}
where $\langle \tilde{\psi}({\bf {r}})\rangle=0$ and $\langle\psi({\bf 
r})\rangle_{t}=\Phi({\bf {r}},t) $ is the macroscopic wavefunction. 
The non-condensate (or excited atom component) field operators  
$\tilde{\psi}({\bf 
{r}})$ and $\tilde{\psi}^{\dag}({\bf {r}})$ satisfy the 
usual Bose commutation relations. Using (3), the 
matrix in (2) splits into two parts
\begin{equation}
\hat{g}(1,1';U)=\hat{\tilde 
{g}}(1,1';U)+\hat{h}(1,1';U).
\end{equation}
Here $\hat{\tilde{g}}$ is identical to (2) except that it involves the 
non-condensate part of the field operators, and
\begin{equation}
\hat{h}(1,1';U)\equiv -i\left ( \begin{array}{cc} 
\Phi(1)\Phi^{*}(1')   \hspace*{5mm} \Phi(1) \Phi(1') \\ 
\Phi^{*}(1)\Phi^{*}(1')\hspace*{5mm} \Phi^{*}(1)\Phi(1') 
\end{array} \right ),
\end{equation}
with $\langle\psi^{\dag}(\r)\rangle_{t}\equiv \Phi^{*}(\r,t)$. In a Bose-condensed 
system, the finite value of $\Phi({\bf r},t)$ leads to finite 
values of the off-diagonal (or anomalous) propagators 
$\langle\tilde{\psi}(1)\tilde{\psi}(1')\rangle$ and $\langle\tilde{\psi}^{\dag}(1)
\tilde{\psi}^{\dag}(1')\rangle$. These must be dealt with on a equal basis 
with the diagonal (or normal) propagators, which is the reason  we must 
work with a $2\times 2$ matrix 
single-particle Green's function.  

A very convenient and elegant way of generating the equations of motion for $\hat{\tilde{g}}$
and $\hat{h}$ is to use functional derivatives with respect to weak 
external fields. The latter are described by 
\begin{equation}
H'(t)=\int d {\bf r} [U( {\bf {r}},t) \tilde{n}( {\bf 
{r}},t)+\eta( {\bf {r}},t) \tilde{\psi}^{\dag}( {\bf 
{r}},t)+\eta^{*} ({\bf {r}},t) \tilde{\psi}( {\bf {r}},t)],
\end{equation}
where $U( {\bf {r}},t) $ is the external generating scalar field, 
while
$\eta( {\bf{r}},t)$ and $\eta^{*} ({\bf {r}},t)$ describe the
 symmetry-breaking fields involving ``particle sources''¥ \cite{hm,bog}.
The higher-order Green's functions can all be expressed as a functional 
derivatives of single-particle Green's functions  
with respect to these fields.

Following the  Kane-Kadanoff (KK) analysis  \cite{kb,kk}, 
the HFB  equations of motion can be conveniently written in the following 
$2 \times 2$ matrix form
\be
 \int d\bar{1}\left[{\hat{g}_{0}}^{-1}(1, 
\bar{1})-\hat{\Sigma}^{HF}¥(1, 
\bar{1})\right] \hat{\tilde{g}}(\bar{1},1') =\delta(1,1'),
\ee

\be
 \int d\bar{1}\hat{\tilde{g}}(1,\bar{1})\left[{\hat{g}_{0}}^{-1} 
(\bar{1},1')-\hat{\Sigma}^{HF}¥( \bar{1},1') \right] 
=\delta(1,1'),
\ee

\be
 \int d\bar{1}\left[ {\hat{g}_{0}}^{-1}(1,\bar{1})-\hat{S}^{HF}(1, 
\bar{1})\right] 
\hat{h}(\bar{1},1')=-i\hat{\eta}^{ext}(1)\langle\hat{\Psi}^{\dag}(1')\rangle,
\ee

\be
 \int d\bar{1}\hat{h}(1,\bar{1})\left[{\hat{g}_{0}}^{-1}(\bar{1},1')-\hat{S}^{HF}
(\bar{1},1') \right]=-i\langle\hat{\Psi}(1)\rangle{\hat{\eta}^{ext \dag}}(1').
\ee
In the above equations, integration over $d{\bar 1}$ means integration 
over the coordinates $({\bf r}_{\bar{1}}, t_{\bar{1}})$,  
$\delta(1,1')\equiv\delta({\bf r}-{\bf r'})\delta(t-t')$, and
we have to introduce spinors
\be
\langle\hat{\Psi}(1)\rangle \equiv \left ( \begin{array}{c}  \Phi(1) \\  
\Phi^{*}¥(1) \end{array}\right ) , \hspace{5mm} 
\langle\hat{\Psi}^{\dag}(1')\rangle \equiv \left(\Phi^{*}¥(1), \Phi(1)\right).
\ee
Here $\hat{\eta}^{ext}$ describes the external particle-source fields defined 
in (6),
\be
\hat{\eta}^{ext}¥(1) =\left ( \begin{array}{c}  \eta(1) \\ 
\eta^{*}(1) \end{array}\right ).
\ee
 The inverse of the
 $2\times 2$ matrix non-interacting propagator ${g}_{0}(1,1')$ is defined by
\be
{g}_{0}^{-1}(1,1')=\left[i{\bf \tau^{(3)}}\frac{\partial}{\partial 
t_{1}}+\frac{\nabla_{1}^{2}}{2m}-U_{ext}({\bf r}_{1})-U(1)+\mu_{0}
\right] \delta(1,1').
\ee
In the HFB approximation, the $2\times 2$ self-energies in (7)-(10) 
are given by \cite{kk,hfb,hm}
\be
\hat{\Sigma}^{HF}¥(1,1')=g\left ( \begin{array}{cc} 
2n(1),  \hspace*{5mm} m(1) \\ 
m^{*}¥(1),\hspace*{5mm} 2n(1) \end{array} \right )\delta(1,1'),
\hspace{5mm}
\hat{S}^{HF}¥(1,1')=g\left ( \begin{array}{cc} 
2\tilde{n}(1)+n_{c}(1),\hspace*{5mm} \tilde{m}(1) \\ 
\tilde{m}^{*}(1),\hspace*{5mm} 2\tilde{n}(1)+n_{c}(1)¥ \end{array} 
\right ) \delta(1,1').
\ee
In the above equations, {\it n}(1) is the {\it total} local density given by
\be
n(1)=i\left[\tilde{g}^{<}_{11}(1,1^{+})+h_{11}(1,1)\right]=\langle\tilde{\psi}
^{\dag}(1)\tilde{\psi}(1)\rangle+\mid \Phi(1)\mid^{2}\equiv \tilde{n}(1)+n_{c}(1),
\ee
 where ${\it \tilde{n}}(1)$ and ${\it n_{c}}(1)$ are non-condensate and 
condensate 
density, respectively. Similarly, ${\it m}(1)$ is the ``anomalous'' local density 
defined by
\ba
&&m(1)=i \left[\tilde{g}_{12}^{<}(1,1)+h_{12}(1,1)\right]
=\langle\tilde{\psi}(1)\tilde{\psi}(1)\rangle+\left[\Phi(1)\right]^{2}=\tilde{m}(1)
+ \left[\Phi(1)\right]^{2}\nonumber \\
&&m^{*}¥(1)=i \left[\tilde{g}_{21}^{<}(1,1)+h_{21}(1,1)\right]
=\langle \tilde{\psi}^{\dag}¥(1)\tilde{\psi}^{\dag}¥(1)\rangle+\left[\Phi^{*}(1)
\right]^{2}=\tilde{m}^{*}(1)+\left[\Phi^{*}(1)\right]^{2}.¥
\ea
These HFB results were first written down in this formalism  by Kane  and 
Kadanoff \cite{kk}, with the ``particle-source''¥ fields  $\eta$ and $\eta^{*}$ 
 left implicit. The equation of motion for the order parameter $\Phi(1)$ 
 in this
  HFB approximation is given by 
\be
\left[i\frac{\partial}{\partial t_{1}}+\frac{\nabla_{1}^{2}}{2m}
+\mu_{0}-U_{ext}(\r_{1})-U(1) -g\left[2\tilde{n}(1)+n_{c}(1)\right]\right]\Phi(1)
=g\tilde{m}(1)\Phi^{*}(1)+\eta(1),
\label{eq:cond}
\ee
and its complex conjugate. Equations (7)-(\ref{eq:cond}) are a closed set of 
equations and define what is called the dynamic HFB approximation.
\newcommand{\op}{\hat{\psi}}
\newcommand{\co}{({\bf R},T)}
\newcommand{\cog}{(\r, {\bf R}, T)}
\newcommand{\cof}{({\bf p}, {\bf R}, T)}
\newcommand{\nee}{\tilde{\epsilon}_{p}}¥

We now turn to solving the HFB equations of motion for atoms in the 
presence of a trapping potential. If the external generating fields induce a 
disturbance with a wavelength much 
longer than thermal wavelengths and frequencies much smaller than 
characteristic particle energies  then the propagator $g(1,1')=g({\bf r},t;{\bf R},T)$ can be expected to vary slowly 
as a function of the center-of-mass coordinates  
\be
{\bf R}=\frac{1}{2}(\r_{1}+\r_{1'}), \hspace{5mm} 
T=\frac{1}{2}(t_{1}+t_{1'}),
\ee
and to be dominated by small values of the relative coordinates
\be
\r=\r_{1}-\r_{1'},\hspace{5mm} 
t=\frac{1}{2}(t_{1}-t_{1'}).
\ee
More precisely, Fourier transforming with respect to {\bf r} and {\it 
t}, the function $g({\bf 
p},\omega;{\bf R},T)$ describes the density of elementary excitations 
of momenta ${\bf p}$ and energy $\omega$ at point $({\bf R},T)$ 
\cite{kb}. These 
quasiparticles are assumed to have high momentum and energy (relative 
to the collective modes we would obtain from the kinetic equations), 
which means that  $g({\bf r},t;{\bf R},T)$ is mainly weighted at small values of the 
coordinates ${\bf r}$ and {\it t}. 

If the Bose-condensate order parameter  $\langle \psi(1)\rangle$ is written in 
terms of amplitude and phase variables
\be
\langle\psi(1)\rangle=[n_{c}(1)]^{\frac{1}{2}}e^{i\theta(1)},
\ee
 we can generalize the usual definitions  for superfluid velocity and 
local chemical potential to  non-equilibrium 
systems by identifying \cite{kk} 
\ba
\nabla_{1}\theta(1)&=&m {\bf v}_{s}(1)     
\nonumber \\
\frac{\partial \theta(1)}{\partial 
t_{1}}&=&-\left[\mu(1)-\mu_{0}+\frac{1}{2}m {v_{s}}^{2}(1)\right]. 
\ea
The superfluid velocity ${\bf v}_{s}$ enters as the gradient of the 
phase of the 
condensate wavefunction and the local chemical potential $\mu(1)$ is 
connected with the time-derivative of the phase.
However, a problem arises in that the phase is a rapidly varying function 
of $({\bf R},T)$ which induces strong 
variations in the off-diagonal elements of $\hat{h}$, and these are 
coupled to the components of  
$\tilde{g}$. To remove this strong $({\bf R},T)$-dependence associated 
with the phase  $\theta$, 
we apply the well-known \cite{kk} gauge transformations on 
$\hat{h}(1,1')$ and $\hat{\tilde{g}}(1,1')$:
\ba
\hat{h'}(1,1')=e^{-i\theta(1){\bf 
\tau^{(3)}}}\hat{h}(1,1')e^{i\theta(1){\bf \tau^{(3)}}} \nonumber \\
\hat{\tilde{g'}}(1,1')=e^{-i\theta(1){\bf 
\tau^{(3)}}}\hat{\tilde{g}}(1,1')
e^{i\theta(1){\bf \tau^{(3)}}}, 
\label{eq:gauge}
\ea
where ${\bf \tau^{(3)}}$ is the Pauli spin matrix.
The physical interpretation of (\ref{eq:gauge}) is that 
it involves a transformation to a coordinate system in which 
non-condensate atoms are moving with average velocity ${\bf v}_{s}$ with respect to 
a stationary condensate. The gauge transformation (\ref{eq:gauge})  removes 
the strong $({\bf R},T)$-dependence associated with the order 
parameter phase $\theta$ and   
leaves the equations of motion (7)-(10) invariant if we replace 
$g_{0}^{-1}$ in (13) by
\be
{g'}_{0}^{-1}(1,1')=\left[ i {\bf \tau^{(3)}}\frac{\partial}{\partial 
t_{1}}-\frac{\partial \theta(1)}{\partial t_{1}}+\frac{1}{2}
[\nabla_{1}+i\nabla_{1}\theta(1) {\bf \tau^{(3)}}]^{2}-U_{ext}({\bf 
r_{1}})-U(1)+\mu_{0} \right] \delta(1,1').
\ee
After carrying out this gauge transformation, the HFB equations for 
$\tilde{g}^{\up}_{11}(1,1';U)$ and $\tilde{g}^{\up}_{12}(1,1';U)$ 
given by (7)-(10) become
\ba
\left[i\frac{\partial}{\partial t_{1}}-\frac{\partial \theta(1)}
{\partial t_{1}}
+\frac{1}{2m}[\nabla_{1}+i\nabla_{1}\theta(1)]^{2}+
\mu_{0}-U_{eff}(1)\right] \tilde{g'}^{\up}_{11}(1,1';U) 
&&=g m'(1)\tilde{g'}^{\up}_{21}(1,1';U) \nonumber \\
\left[ i\frac{\partial}{\partial t_{1}}-\frac{\partial \theta(1)}
{\partial t_{1}}+\frac{1}{2m}[\nabla_{1}+i\nabla_{1}\theta(1)]^{2}
+ \mu_{0}-U_{eff}(1)\right]\tilde{g'}^{\up}_{12}(1,1';U) 
&&=g m'(1)\tilde{g'}^{\up}¥_{22}(1,1';U) \nonumber \\
\left[-i\frac{\partial}{\partial t_{1'}}-\frac{\partial \theta(1')}
{\partial t_{1'}}+\frac{1}{2m}[\nabla_{1'}-i\nabla_{1'}\theta(1')]^{2}
+ \mu_{0}-U_{eff}(1')\right] \tilde{g'}^{\up}_{11}(1,1';U) 
&&=g m'^{*}(1')\tilde{g'}^{\up}¥_{12}(1,1';U) \nonumber \\
\left[ i\frac{\partial}{\partial t_{1'}}-\frac{\partial \theta(1')}
{\partial t_{1'}}+\frac{1}{2m}[\nabla_{1'}+i\nabla_{1'}\theta(1')]^{2}
+ \mu_{0}-U_{eff}(1')\right]\tilde{g'}^{\up}_{12}(1,1';U) 
&&=g m'(1')\tilde{g'}^{\up}¥_{11}(1,1';U). 
\label{eq:movingg}
\ea
Here 
\be
U_{eff}(1)\equiv U_{ext}(\r_{1})+U(1)+2gn'(1) \label{eq:ueff}
\ee
is the effective self-consistent Hartree-Fock dynamic mean field. 
The condensate part $2gn_{c}(1)$ in (\ref{eq:ueff}) can be viewed as an 
additional ``external field''¥ acting on the non-condensate. 
Since we will work with these gauge-transformed correlation functions 
in the rest of this paper, we drop the primes 
on $\tilde{g}_{11}$, $\tilde{g}_{12}$, {\it n} and {\it m} to simplify the notation.

The corresponding equation of motion for the condensate amplitude 
$\sqrt{n_{c}(1)}$ is in the 
moving frame of reference can be written in the form:
\ba
&&\left[i\frac{\partial}{\partial t_{1}}-\frac{\partial 
\theta(1)}{\partial 
t_{1}}+\frac{\nabla_{1}^{2}}{2m}-\frac{1}{2}m{\bf v}_{s}^{2}+\right.
\mu_{0}¥-U_{ext}(\r_{1})-U(1) -g\left[ 2\tilde{n}(1)+n_{c}(1)\right]+ 
 \nonumber \\
&&+i{\bf v}_{s}(1)\cdot \nabla_{1}+\left.
\frac{i}{2}\nabla_{1}\cdot{\bf v}_{s}(1)\right]\sqrt{n_{c}(1)}
=g\tilde{m}(1)\sqrt{n_{c}(1)}+\eta'(1),
\label{eq:amplitude1}
\ea
where $\eta'(1)\equiv \eta(1)e^{-i\theta(1)}¥$ is the 
symmetry-breaking source function in 
the moving frame of reference. 
We note that there is a factor of two  difference between how the 
condensate and non-condensate enter these equations. This is 
because atoms in the condensate are in the same state and thus 
there is no exchange part. In the case of non-condensate atoms,  both 
Hartree and  Fock terms arise since we are dealing with atoms in 
different states. 

\section{HFB kinetic equations}
To rewrite the equations of motion derived in Section II in the form of kinetic equations, we recall the connection between the 
usual single-particle distribution function $f_{1}¥({\bf p},{\bf R},T)$ and 
the diagonal Green's function $\tilde{g}^{<}_{\alpha 
\alpha}(1,1')$. We define (see p. 67 of Ref. [1]):
\ba
f_{1}({\bf p},{\bf R},T)&&\equiv \int d\r e^{-i{\bf p}\cdot 
\r}\left[i \tilde{g}^{<}_{11}({\bf r},t=0;{\bf R},T)\right] \nonumber \\
&&=\int d\r e^{-i{\bf p}\cdot 
\r}\langle \tilde{\psi}_{U}¥^{\dag}({\bf R}-\frac{\r}{2},T)\tilde{\psi}_{U}
({\bf R}+\frac{\r}{2},T)\rangle \label{eq:f1}
\ea
where, by definition,
\be 
\int \frac{d{\bf p}}{(2\pi)^{3}}f_{1}({\bf p},{\bf R}, 
T)=\tilde{n}({\bf R},T).
\ee
We can see that $f_{1}({\bf p},{\bf R},T)$ corresponds to the 
well-known Wigner distribution function. In the classical limit, 
it reduces to the distribution function giving the number of atoms 
 with momentum ${\bf p}$ at point ${\bf R}$ and time {\it T}. 
The symmetry-breaking terms in (6) result in the finite 
value of anomalous Green's function, and thus it is natural to 
introduce an additional distribution function for the non-condensate atoms
which will give us the anomalous non-condensate density ${\it 
\tilde{m}}(1)$, namely
\ba
f_{2}({\bf p},{\bf R},T)&&\equiv \int d\r e^{-i{\bf p}\cdot 
\r}\left[i \tilde{g}^{<}_{12}({\bf r},t=0;{\bf R},T)\right] \nonumber \\
&&=\int d\r e^{-i{\bf p}\cdot 
\r}\langle{\tilde{\psi}}_{U}({\bf R}-\frac{\r}{2},T){\tilde{\psi}}_{U}
({\bf R}+\frac{\r}{2},T)\rangle. \label{eq:f2}
\ea
One can easily verify that the pair correlation function $\tilde{m}$ 
in (16) is given by
\be 
\int \frac{d{\bf p}}{(2 \pi)^{3}}f_{2}({\bf p},{\bf R},T)
=\tilde{m}({\bf R},T).
\ee

It is important to remember that, as defined, the distribution 
functions $f_{1}$ and $f_{2}$ describe the behaviour of the atoms. 
They should not be confused with the distribution function for the 
quasiparticle excitations, such as discussed in Ref.\cite{kd}.

To obtain  kinetic equations, we
follow the Kadanoff-Baym approach in \cite{kb,kk} and rewrite the equations for 
$\tilde{g}_{11}^{<}¥(1,1')$ and $\tilde{g}_{12}^{<}¥(1,1')$
in the relative and center-of-mass coordinates $({\bf r},t;{\bf R},T)$. 
We could obtain equations for $g({\bf r},t;{\bf R},T)$ as in 
Ref.\cite{kk}, but for the simple HFB self-energies of 
interest it is sufficient to consider {\it t}=0, {\it i.e.}  set
$t_{1'}=t_{1}^{+}=T$ (see Ch. 7 of \cite{kb}). Using (\ref{eq:movingg}), this 
procedure gives
\ba
&&\left[ i\frac {\partial}{\partial T}+\Delta U^{\mu}_{eff}({\bf r}, 
{\bf R},T)+\frac{1}{m}\nabla_{{\bf r}}\cdot \right.
\nabla_{\bf {R}}+i {\bf v}_{s}^{-}({\bf r},{\bf R},T) 
\cdot \nabla_{\bf r}+ \frac{i}{2}{\bf v}_{s}^{+}({\bf r},{\bf R},T) 
\cdot \nabla_{{\bf R}} \nonumber \\
&&+\left. \frac{i}{2}\nabla_{\bf r} \cdot {\bf v}_{s}^{-}
({\bf r},{\bf R},T)+ \frac{i}{4}\nabla_{\bf {R}} \cdot {\bf 
v}_{s}^{+}({\bf r},{\bf R},T) \right] \tilde{g}_{11}^{<}({\bf r},
{\bf R},T)  \nonumber \\
&&=g \left[ m({\bf R}+\frac{\bf r}{2},T)\tilde{g}_{21}^{<}({\bf r},{\bf R},T)-
m^{*}({\bf R}-\frac{\bf r}{2},T)\tilde{g}_{12}^{<}({\bf r},{\bf 
R},T)\right], \label{eq:f1exact} 
\ea
where we have introduced the abbreviations
\be
\Delta U^{\mu}_{eff}(1,1')\equiv 
\mu(1)-\mu(1')-U_{eff}(1)+U_{eff}(1'),
\ee
and
\be
{\bf v}_{s}^{\pm}({\bf r},{\bf R},T)\equiv {\bf v}_{s}({\bf R}
 +\frac{{\bf r}}{2},T) \pm {\bf v}_{s}({\bf R} -\frac{{\bf r}}{2},T) .
\ee
Similarly, the equation of motion for $\tilde{g}_{12}^{<}¥(1,1')$¥ expressed in the
  $({\bf r},{\bf R};t,T)$ variables gives
\ba
&&\left[ i\frac {\partial}{\partial T}+\mu({\bf R}+\frac{\r}{2},T)+
\mu({\bf R}-\frac{\r}{2},T)-U_{eff}¥({\bf R}+\frac{\r}{2},T)
-U_{eff}¥({\bf R}-\frac{\r}{2},T)+\frac{\nabla_{\bf 
R}^{2}}{4m}+\right. 
\nonumber \\
&&+\frac{1}{m}\nabla_{\r}^{2}+i {\bf v}_{s}^{-}({\bf r},{\bf R},T) 
\cdot \nabla_{\bf r}+ \frac{i}{2}{\bf v}_{s}^{+}({\bf r},{\bf R},T) 
\cdot \nabla_{{\bf R}} 
+ \frac{i}{2}\nabla_{\bf r} \cdot {\bf v}_{s}^{-}
({\bf r},{\bf R},T)+ \nonumber \\
&& \left. +\frac{i}{4}\nabla_{\bf {R}} \cdot {\bf 
v}_{s}^{+}({\bf r},{\bf R},T) \right] \tilde{g}_{12}^{<}({\bf r},
{\bf R},T)  
=g \left[m({\bf R}+\frac{\bf r}{2},T)\tilde{g}_{22}^{<}({\bf r},{\bf R},T)+
m({\bf R}-\frac{\bf r}{2},T)\tilde{g}_{11}^{<}({\bf r},{\bf R},T)\right]. 
\label{eq:f2exact}
\ea
Finally, the equation of motion (\ref{eq:amplitude1}) for the {\it amplitude} of the order 
parameter can be written in the $({\bf R},T)$ coordinates as
\ba
&&\left[ i\frac{\partial}{\partial T}-\frac{\partial \theta({\bf 
R},T)}{\partial T}+
\frac{\nabla_{\bf R}^{2}}{2m}-\frac{1}{2}m{\bf v}_{s}^{2}({\bf R},T) 
+\mu_{0}¥-U_{ext}({\bf R})-U({\bf R},T) \right. \nonumber  \\
&&\left.-g\left[2\tilde{n}({\bf R},T)+n_{c}({\bf R},T)\right]+i{\bf v}_{s}({\bf 
R},T)\cdot 
\nabla_{{\bf R}}+\frac{i}{2}\nabla_{{\bf R}}\cdot{\bf 
v}_{s}({\bf R},T)\right]\sqrt{n_{c}({\bf R},T)} \nonumber \\
&&=g\tilde{m}({\bf R},T)\sqrt{n_{c}({\bf R},T)}+\eta'({\bf R},T).
\label{eq:amplitudeRT}
\ea
\hspace*{10mm}The coupled set of HFB equations given by 
(\ref{eq:f1exact}),(\ref{eq:f2exact}) and 
(\ref{eq:amplitudeRT})
are the main formal results of this paper. These are a straightforward 
generalization of the analogous equations in the normal Hartree-Fock 
approximation discussed by Kadanoff and Baym (see Eq. (7.7) on pg. 71 
of Ref.\cite{kb}). These results are 
important since they allow us to go beyond the simple HFP approximation 
which has been the basis of recent work on the non-equilibrium 
properties of a trapped Bose-condensed gas (see, for example, \cite{zgn,stoof}). 

We next proceed to use (\ref{eq:f1exact}) and (\ref{eq:f2exact}) to 
derive self-consistent equations for $f_{1}({\bf p},{\bf R},T)$ and 
$f_{2}({\bf p},{\bf R},T)$ for the case when 
the external perturbation varies slowly in space and time. 
In this case, we expect that physical quantities ${\bf 
v}_{s}$, $\mu$, U, $U_{ext}$, {\it n}(1), and {\it m}(1) all vary 
slowly as functions of the center-of-mass coordinates $({\bf R},T)$.
Thus, in the lowest approximation, and using the fact that small 
values of {\bf r} are most important, we can use
\be
{\bf v}_{s}({\bf R} \pm \frac{{\bf r}}{2},T)={\bf v}_{s}
({\bf R},T) \pm  \left[\frac{{\bf r}}{2}\cdot 
\nabla_{\bf {R}}\right]{\bf v}_{s}({\bf R},T) \nonumber ,
\ee
and hence
\ba
&&{\bf v}_{s}^{+}¥({\bf r},{\bf R},T)\simeq 2{\bf v}_{s}({\bf R},T), \hspace{10mm}
{\bf v}_{s}^{-}¥({\bf r},{\bf R},T)\simeq \left[{\bf r}\cdot \nabla_{\bf 
R}\right]{\bf v}_{s}({\bf R},T), \nonumber \\
&&\Delta U^{\mu}_{eff}({\bf r},{\bf R},T) \simeq -{\bf 
r}\cdot \nabla_{{\bf R}}[U_{eff}({\bf R},T)-\mu({\bf R},T)].
\ea
If we rewrite (\ref{eq:f1exact}) using (\ref{eq:f1}) and (\ref{eq:f2}), and Fourier transform 
it, we obtain after some algebra
\ba
\left[ \frac{\partial}{\partial T}\right.&& -\left.\nabla_{\bf {
R}}[\nee+{\bf v}_{s}\cdot {\bf p}]\cdot \nabla_{{\bf 
p}}+\nabla_{\bf{ 
p}}[\nee+{\bf v}_{s}\cdot {\bf p}]\cdot \nabla_{\bf{ R}} 
\right] f_{1}({\bf p},{\bf R},T) \nonumber \\
&&=-ig\left[ m({\bf R},T)f_{2}(-{\bf 
p},{\bf R},T)-m^{*}({\bf R},T)f_{2}({\bf p},{\bf R},T) \right] 
\nonumber \\
&&+\frac{g}{2}\left[\nabla_{\bf R}m({\bf R},T)\cdot \nabla_{\bf 
p}f_{2}(-{\bf p},{\bf R},T)+\nabla_{\bf R}m^{*}({\bf R},T)\cdot 
\nabla_{\bf p}f_{2}({\bf p},{\bf R},T)\right],
\label{eq:slowf1}
\ea
where the ``normal''¥ single-particle  energy is defined by
\be
\nee({\bf R},T)\equiv \frac{p^{2}}{2m}+U_{eff}({\bf R},T)-\mu({\bf R},T).
\label{eq:energy}
\ee
We emphasize that $\nee$ is not, in general, the local HFB excitation energy.
The corresponding kinetic equation for $f_{2}({\bf p},{\bf R},T)$ is:
\ba
\left[\frac{\partial}{\partial T}\right.&&+\left.i2\nee({\bf 
R},T)-\nabla_{\bf R}\left[{\bf v}_{s}\cdot {\bf p}\right]\cdot 
\nabla_{\bf p}+ \nabla_{\bf p}\left[{\bf v}_{s}\cdot {\bf p}\right]\cdot 
\nabla_{\bf R} \right]
f_{2}({\bf p},{\bf R},T)= \nonumber \\
&&-igm({\bf R},T)\left[ f_{1}({\bf p},{\bf 
R},T) +f_{1}(-{\bf p},{\bf R},T) +1 \right] \nonumber \\
&&+\frac{g}{2}\nabla_{\bf R}m({\bf R},T)\cdot\nabla_{\bf p}¥\left[f_{1}({\bf 
p},{\bf R},T)-f_{1}(-{\bf p},{\bf R},T)\right].
\label{eq:slowf2}
\ea
These are the coupled HFB collisionless kinetic equations in a frame 
moving with the velocity ${\bf v}_{s}$, for 
the case of slowly varying disturbances.

The equation of motion (\ref{eq:amplitudeRT}) for the amplitude of the order 
parameter is an exact equation, and is not limited for a slowly varying 
disturbances. Equating the real and imaginary parts of 
(\ref{eq:amplitudeRT}), we obtain two hydrodynamic equations of motion for the 
condensate 
\ba
&&\frac{\partial n_{c}({\bf R},T)¥}{\partial T}=-\nabla_{\bf 
R}\left[n_{c}({\bf R},T){\bf v}_{s}({\bf R},T) 
\right]+2gn_{c}({\bf R},T)Im\left[\tilde{m}({\bf R},T)\right],\label{eq:timenc} 
\\
&&\frac{\partial \theta({\bf R},T)¥}{\partial T}+\frac{1}{2}m{\bf 
v}_{s}^{2}
({\bf R},T)-\mu_{0}=-\mu({\bf R},T)-\eta'({\bf 
R},T)\frac{1}{\sqrt{n_{c}({\bf R},T)}} \label{eq:timetheta},
\ea
where the condensate chemical potential $\mu({\bf R},T)$ is defined by
\be
\mu¥({\bf R},T)\equiv -\frac{\nabla_{\bf R}^{2}\sqrt{n_{c}({\bf 
r},T)}}{2m\sqrt{n_{c}({\bf R},T)}}+U_{ext}({\bf R})+U({\bf R},T)+
g\left[2\tilde{n}({\bf R}, 
T)+n_{c}({\bf R}, T)\right]+g Re\left[\tilde{m}({\bf R}, T)\right].
\label{eq:cmpotential}
\ee
Taking the gradient of (\ref{eq:timetheta}), we obtain the generalized Landau 
equation
 \cite{zgn} 
\be
m\left[ \frac{\partial {\bf v}_{s}({\bf R},T)}{\partial T} 
+\frac{1}{2}\nabla_{\bf R}{\bf v}_{s}^{2}({\bf 
R},T)\right]=-\nabla_{\bf R}\left[¥\mu({\bf R},T)+\eta'({\bf 
R},T)\frac{1}
{\sqrt{n_{c}({\bf R},T)}}\right].
\label{eq:gradvelocity}
\ee
In ZGN \cite{zgn}, the contribution of the external symmetry-breaking 
field $\eta'$ was left implicit. 
In the often used Thomas-Fermi (TF) approximation, the kinetic energy of the 
condensate is omitted and the HFB 
approximation for the condensate chemical potential then simplifies to
\be
\mu^{TF}¥({\bf R},T)=U_{ext}({\bf R})+2g\tilde{n}({\bf R},T)+gn_{c}({\bf 
R},T)+gRe\left[\tilde{m}({\bf R},T)\right].
\label{eq:tfmu}
\ee 

Next we consider Eq.(\ref{eq:slowf1}) in the high temperature limit. In that 
case, $n_{c}$ is small and therefore we can neglect the terms 
proportional to $gn_{c}f_{2}$ in (\ref{eq:slowf1}) as small. Also, we know that 
$\tilde{m}$ must be at least of order {\it g} \cite{varenna}, and 
therefore the $g\tilde{m}$ contribution to the self-energy is ${\cal O}(g^{2})$.
Thus, we can neglect the right side of (\ref{eq:slowf1}), leaving 
\be
\left[ \frac{\partial}{\partial T} -\nabla_{\bf {
R}}[\nee+{\bf v}_{s}\cdot {\bf p}]\cdot \nabla_{{\bf 
p}}+\nabla_{\bf{ 
p}}[\nee+{\bf v}_{s}\cdot {\bf p}]\cdot \nabla_{\bf{ R}} 
\right] f_{1}({\bf p},{\bf R},T) =0. 
\label{eq:hightf1}
\ee
This is precisely the expected collisionless Boltzmann equation for 
$f_{1}$, valid in the HFP approximation. This approximation 
is only valid at finite temperatures, close to $T_{BEC}$, in which case 
$\nee({\bf R},T)$ as defined in (\ref{eq:energy}) {\it is} the correct excitation energy.
In this limit, the kinetic equation (\ref{eq:hightf1}) becomes 
equivalent to that derived (using a different formalism) by Kirkpatrick and Dorfman (KD) 
\cite{kd} for a uniform gas. It is the local rest frame equivalent of 
the kinetic equation used by ZGN \cite{zgn} if one ignores the collision terms. 

\section{Static HFB equilibrium solutions in the semi-classical 
approximation}

For a uniform system, the matrix Green's functions defined in (4) depend only on 
the relative  coordinates, i.e. $\tilde{g}_{\alpha \beta}(1,1')=
\tilde{g}_{\alpha \beta}(1-1')$. One can then solve (7)-(10) by 
Fourier transformation to obtain expressions for 
$\tilde{g}_{11}({\bf p},\omega)$ and $\tilde{g}_{12}({\bf 
p},\omega)$. Using these, we obtain the single-particle spectral 
density in the form \cite{fw}
\ba
A_{11}({\bf p},\omega)&&=-2 Im\tilde{g}_{11}({\bf p},\omega+i0^{+})= -2 Im\left[
\frac{u_{p}^{2}}{\omega-E_{p}+i0^{+}}-\frac{v_{p}^{2}}{\omega+E_{p}+i0^{+}}\right]
\nonumber \\
&&=2\pi\left[u_{p}^{2}\delta(\omega-E_{p})- 
v_{p}^{2}\delta(\omega+E_{p})\right] \nonumber \\
A_{12}({\bf p},\omega)&&=-2Im\tilde{g}_{12}({\bf p},\omega+i0^{+}¥)=2Im\left[
\frac{u_{p}v_{p}^{*}}{\omega-E_{ p}+i0^{+}}-\frac{u_{p}v_{p}^{*}}{\omega+E_{ p}+i0^{+}}\right]
\nonumber \\
&&=-2\pi u_{p}v_{p}^{*}\left[\delta(\omega-E_{p})
-\delta(\omega+E_{p})\right].
\label{eq:unspectral}
\ea
The HFB excitation energy $E_{p}$ is given by¥
\be
E_{p}^{2}¥=\left[\frac{p^{2}}{2m}+2gn-\mu \right]^{2}¥-(gm)^{2},
\label{eq:unenergy}
\ee
with
\be
\nee\equiv\frac{p^{2}}{2m}+2gn-\mu,
\ee
and
\be
u_{p}^{2}=\frac{1}{2}\left[\frac{\nee}{E_{p}}+1\right],
\hspace{5mm}
v_{p}^{2}=\frac{1}{2}\left[\frac{\nee}{E_{p}}-1\right],
\hspace{5mm}
u_{p}v_{p}^{*}=\frac{gm}{2E_{p}}.
\ee
The results in (\ref{eq:unspectral}) have the same structure as in the simpler 
Bogoliubov approximation (see Ch.14 of Ref \cite{fw}). 
From (\ref{eq:cond}), with $\Phi(1)=const.$, it follows that chemical potential in 
HFB approximation is given by \cite{hfb,hm}
\be
\mu=g\left(n+\tilde{n}+\tilde{m}\right).
\ee
If we use this result in (\ref{eq:unenergy}), it reduces to
\be
E_{p=0}^{2}=g^{2}\left[n_{c}-\tilde{m}\right]^{2}-g^{2}\left[n_{c}+
\tilde{m}\right]^{2}=-4g^{2}\tilde{m}n_{c}.
\ee
Therefore, the long-wavelength HFB excitation spectrum has a finite energy gap.
 
The density of non-condensate atoms in a uniform system can be found using the 
single-particle spectral density given in (\ref{eq:unspectral}) 
\ba
\tilde{n}&=&\int \frac{d{\bf p}}{(2\pi)^{3}}\int \frac{d\omega}{2\pi} 
N_{0}(\omega)A_{11}¥({\bf 
p},\omega) \nonumber \\ 
&=&\int \frac{d{\bf p}}{(2\pi)^{3}}
\left[v_{p}^{2}+\left(u_{p}^{2}+v_{p}^{2}\right)N_{0}(E_{p})¥\right] 
\nonumber\\
&=&\int \frac{d{\bf 
p}}{(2\pi)^{3}}\left[\frac{\tilde{\varepsilon}_{p}}{2E_{p}}\left[2N_{0}(E_{p})
+1\right]-\frac{1}{2}\right],
\label{eq:untilden}
\ea
where $N_{0}(E_{p}¥)$ is the Bose distribution function. Similarly, 
the anomalous density is given by
\ba
\tilde{m}&=&\int \frac{d{\bf p}}{(2\pi)^{3}}\int \frac{d\omega}{2\pi}
N_{0}(\omega)A_{12}¥({\bf 
p},\omega) \nonumber \\ 
&=&-\int \frac{d{\bf p}}{(2\pi)^{3}}
v_{p}u_{p}^{*}¥\left[2N_{0}(E_{p})+1\right]\nonumber \\
&=&-\int \frac{d{\bf 
p}}{(2\pi)^{3}}\frac{gm}{2E_{p}}\left[2N_{0}(E_{p})+1\right].
\label{eq:untildem}
\ea

We want to find an approximate solution for $f_{1}$ and $f_{2}$ for a 
trapped Bose gas which will be
 the analogue of the uniform gas results in (\ref{eq:untilden}) and 
(\ref{eq:untildem}).   
We start from the coupled static HFB equations, as derived in Ref.\cite{hfb}:
\ba
\hat{\cal L}u_{i}({\bf R})-gm_{0}({\bf R})v_{i}({\bf R})&=&E_{i}u_{i}({\bf 
R}) \nonumber \\
\hat{\cal L}v_{i}({\bf R})-gm_{0}({\bf R})u_{i}({\bf R})&=&-E_{i}v_{i}({\bf 
R}), 
\label{eq:statichfb}
\ea
where
\be
\hat{\cal L}=-\frac{\nabla^{2}¥}{2m}+U_{ext}({\bf R})-\mu+2gn({\bf R}).
\ee
We can solve (\ref{eq:statichfb}) using the well-known 
semi-classical approximation. We assume that the normal and anomalous 
density are smooth functions of {\bf R} on the scale of length of 
$a_{HO}\equiv \sqrt{\frac{\hbar}{m\omega_{0}}}$, which defines the 
size of the condensate in a harmonic potential with a trap frequency 
$\omega_{0}$ (we reinsert $\hbar$ in this discussion for physical 
clarity). Hence 
the self-consistent Hartree-Fock mean field varies slowly (as a function 
of {\bf R}) on the length scale of order $a_{HO}$. Therefore, we can 
assume that $u_{i}({\bf R})$ and $v_{i}({\bf R})$ have a form of a 
plane waves with a slowly varying amplitude in that region 
\cite{stringari}, i.e.
\ba
&&u_{i}({\bf R})\equiv u({\bf p},{\bf R})e^{i\varphi({\bf R})}
\nonumber \\
&&v_{i}({\bf R})\equiv v({\bf p},{\bf R})e^{i\varphi({\bf R})}.
\label{eq:planewaves}
\ea
We introduce the momentum of elementary excitations by
\be
{\bf p}=\hbar \nabla_{\bf R}\varphi({\bf R}),
\ee
which satisfy the condition ${\bf p}\gg \hbar/a_{HO}$. This 
condition, when expressed in terms of wavelength reduces to the small 
wavelength limit ($\lambda \ll a_{H0}$), or equivalently, to the 
semi-classical approximation limit expressed as $kT \gg \hbar 
\omega_{0}¥$, where $\hbar \omega_{0}$ gives the harmonic well energy 
level spacing. If 
this condition is satisfied, we can neglect the spatial derivatives of {\it u} 
and {\it v}, as well as the second spatial derivative of $\varphi$. This is consistent 
with a so-called
{\it quasi-classical condition}, which requires that a spatial change in wavelength 
of the particle must satisfy the condition $d\lambda/dx \ll 1$ \cite {landau}. The assumed form given by  
(\ref{eq:planewaves}) is only valid in the regions of 
space where this condition is satisfied. To treat the condensate in 
the corresponding approximation, we use the Thomas-Fermi 
approximation which is valid in the large N limit. The only region where 
the TF approximation for the order parameter is inadequate is close to the classical turning 
points at the condensate boundary \cite{stringari,strtheory}, which is consistent with inapplicability of 
the semi-classical approximation near these points.

Putting all this together,
we can easily solve the coupled equations (\ref{eq:statichfb}) for 
$u({\bf p},{\bf R})$ and $v({\bf p},{\bf R})$, 
\be
u^{2}({\bf p},{\bf R})=\frac{\nee({\bf R})+E_{p}({\bf R})}{2E_{p}({\bf R})}, 
\hspace{3mm} 
v^{2}({\bf p},{\bf R})=\frac{\nee({\bf R})-E_{p}({\bf R})}{2E_{p}({\bf R})},
\hspace{3mm}
u({\bf p},{\bf R})v^{*}¥({\bf p},{\bf R})=\frac{gm_{0}{\bf R})}{2E_{p}({\bf R})}.
\ee
The local HFB quasiparticle energy $E_{p}({\bf R})$  is given by
\be
E_{p}^{2}¥({\bf R})=\nee^{2}({\bf R})-(gm_{0}¥({\bf R}))^{2}=\left[\frac{p^{2}}{2m}-\mu_{0}+U_{ext}({\bf R})+2gn_{0}({\bf R})\right]^{2}¥
-(gm_{0}¥({\bf R}))^{2}.
\label{eq:semienergy}
\ee
We use the expressions for $\tilde{n}$ and $\tilde{m}$ 
given in terms of {\it u} and {\it v} \cite{hfb}
\ba
&&\tilde{n}({\bf R})=\sum_{i}\left(\left[\mid u_{i}({\bf R})\mid^2
+\mid v_{i}({\bf R})\mid^2\right]N_{0}(E_{i})+\mid v_{i}({\bf R})\mid^2 \right),
 \nonumber \\
&&\tilde{m}({\bf R})=-\sum_{i}u_{i}({\bf R})v_{i}^{*}({\bf R})\left[2N_{0}(E_{i})+1\right],
\ea
where $N_{0}(E)$ is the Bose distribution for the quasiparticle excitations.
The sum over 
the quantum states is replaced  by the integral $\int d{\bf 
p}/(2\pi)^{3}$ and the semi-classical approximation for the  
diagonal distribution function $f_{10}({\bf p},{\bf R})$ is given by 
\ba
f_{10}({\bf p},{\bf R})&=&v^{2}({\bf p},{\bf R})+\left[u^{2}({\bf p},{\bf 
R})
+v^{2}({\bf p},{\bf R})\right]N_{0}(E_{p}({\bf R}))\nonumber \\&=&
\frac{\nee({\bf R})}{2E_{p}({\bf 
R})}\left[ 2N_{0}(E_{p}({\bf R}))+1\right]-\frac{1}{2},
\label{eq:semif1}
\ea
while for the off-diagonal distribution function $f_{20}({\bf p},{\bf 
R})$ we obtain¥
\be
f_{20}({\bf p},{\bf R})=-\frac{gm_{0}({\bf R})}{2E_{p}({\bf 
R})}\left[2 N_{0}(E_{p}({\bf R}))+1\right].
\label{eq:semif2}
\ee

In summary, in the semi-classical approximation, the local static equilibrium normal density 
is given by
\be
\tilde{n}_{0}¥({\bf R})=\int\frac{d{\bf p}}{(2\pi)^{3}}\left[\frac{\nee({\bf R})}{2E_{p}({\bf 
R})}\left[ 2N_{0}(E_{p}({\bf R})+1\right]-\frac{1}{2}\right],
\label{eq:semin}
\ee
while the local static equilibrium anomalous density by
\be
\tilde{m}_{0}¥({\bf R})=-gm_{0}({\bf R})\int \frac{d{\bf p}}{(2\pi)^{3}}
\left[\frac{2 N_{0}(E_{p}({\bf R}))+1}{2E_{p}({\bf R})}\right].
\label{eq:semim}
\ee
As expected, these semi-classical approximation  
 results are the natural generalizations of the results obtained 
for a {\it uniform} gas given by (\ref{eq:untilden}) and (\ref{eq:untildem}).  
The same kind of semi-classical results have also been obtained in 
Ref. \cite{stringari} for a trapped Bose gas using the Popov 
approximation (which corresponds to the HFB with $\tilde{m}$=0). 
Since the local quasiparticle energy $E_{p}({\bf R})$ given by 
(\ref{eq:semienergy}) depends on 
the normal and anomalous densities, the quantities in (\ref{eq:semienergy}), 
(\ref{eq:semin}) and (\ref{eq:semim}) must be 
solved self-consistently, as in Ref.\cite{stringari}. 

We can now show that these semi-classical static HFB results for 
$f_{10}$ and $f_{20}$ satisfy 
our collisionless static HFB kinetic equations. First of all, we note 
that in the static limit, (\ref{eq:slowf2}) reduces to 
\ba
2\nee({\bf R}) f_{20}({\bf p},{\bf R}) 
=&&-g m_{0}({\bf R}) \left[f_{10}({\bf p},{\bf R}) +f_{10}(-{\bf p},
{\bf R})+1\right]\nonumber \\
&&-i\frac{g}{2}\nabla_{\bf R}m({\bf R},T)\cdot \nabla{\bf p}\left[f_{10}({\bf p},{\bf R})-f_{10}(-{\bf p},{\bf R})\right].
\label{eq:staticf2}
\ea
In  static equilibrium without any mass current, we  have $f_{10}({\bf p},{\bf R})=f_{10}(-{\bf 
p},{\bf R})$, and thus  (\ref{eq:staticf2}) can be further simplified to 
\be
f_{20}({\bf p},{\bf R})=-\frac{g m_{0}({\bf R})}{2\nee({\bf R})} 
\left[2f_{10}({\bf p},{\bf R}) +1\right].
\label{eq:f20}
\ee
We note that the HFB approximation for $\nee({\bf R})$ in (\ref{eq:energy}) is given by
\be
\nee({\bf R})=\frac{p^{2}}{2m}+\frac{\nabla^{2}¥_{\bf 
R}\sqrt{n_{c0}({\bf R})}}{2m\sqrt{n_{c0}({\bf R})}}+g\left[n_{c0}({\bf 
R})-\tilde{m}_{0}({\bf R})\right].
\ee
Similarly, in static thermal equilibrium, Eq.(\ref{eq:slowf1}) for the diagonal
 distribution function $f_{1}\cof$ reduces to 
\be  
\left[ -\nabla_{\bf R}\nee({\bf R}) \cdot \nabla_{\bf p}+
\nabla_{\bf p}\nee({\bf R}) \cdot \nabla_{\bf R}
\right] f_{10}({\bf p},{\bf R}) =g\nabla_{\bf R}m_{0}({\bf R})\cdot\nabla_{\bf 
p}f_{20}({\bf p},{\bf R}).
\label{eq:staticf1}
\ee
In the last step, we again have used that $f_{20}(-{\bf p},{\bf R})=f_{20}({\bf p},{\bf 
R})$ and that $m_{0}$ is real.

The results discussed above for  $f_{10}({\bf p},{\bf R})$ 
and  $f_{20}({\bf p},{\bf R})$ describe local thermal equilibrium distributions
 functions induced by collisions, which are 
not included in our HFB collisionless equations. However, these local 
equilibrium functions do satisfy the static HFB kinetic equations.
Inserting (\ref{eq:semif1}) into (\ref{eq:staticf2}), we obtain 
(\ref{eq:semif2}). This shows that the static equilibrium kinetic equation 
(\ref{eq:f20}) for the off-diagonal distribution function has a 
 solution consistent with the semi-classical approximation.  
Substituting the
local equilibrium distribution functions $f_{10}$ and $f_{20}$ given by
 (\ref{eq:semif1}) and (\ref{eq:semif2}) , a lengthy but 
 straightforward calculation shows that they satisfy 
the static HFB kinetic equation (\ref{eq:staticf1}). 

We recall that the semi-classical local equilibrium expressions for 
$f_{1}$ and $f_{2}$ are only valid under the condition $k_{B}T\gg 
\hbar \omega_{0}$ and $E_{\bf p}({\bf R})\gg \hbar \omega_{0}$. This 
means that these approximate forms cease to be valid at very low 
temperatures (for further discussion, see Ref.\cite{stringari}).
\newcommand{\car}{{\bf R}}
\newcommand{\p}{{\bf p}}
\newcommand{\ta}{\mbox{\boldmath $\eta$}}

\section{Kohn mode}
In this section, following the approach used in Ref.\cite{gz}, we prove
 that our coupled 
equations exhibit the rigid in-phase oscillation (or Kohn mode) 
\cite{zgn}.
This rigid in-phase center-of-mass oscillation of the 
condensate and the non-condensate corresponds to
\ba
&& n_{c}(\car,T) \equiv n_{c0}(\car-\ta(T)) \nonumber \\
&& \tilde{n}(\car,T)\equiv \tilde{n}_{0}(\car- \ta (T)) \nonumber \\
&& \tilde{m}(\car,T) \equiv \tilde{m}_{0}¥(\car-\ta(T)).
\label{eq:kohnde}
\ea
Here the center-of-mass displacement $\ta(T)$ (with ${\bf v}_{s}={\bf 
v}_{n}=\dot{\ta}$)
is independent of position and satisfies the harmonic oscillator
equation of motion
\be
m\frac{\partial^{2}\eta_{\alpha}¥}{\partial T^{2}}=-\omega_{\alpha}^{2}
¥\eta_{\alpha},
\label{eq:kohnosc}
\ee
where $\omega_{\alpha}$ is the harmonic well trap frequency in the $\alpha^{th}$ 
direction. 

We recall equation of motion  for the superfluid velocity given by 
(\ref{eq:gradvelocity}). For the case of a rigid in-phase oscillation 
described by (\ref{eq:kohnde}), this equation can be written in the form
\be
m\frac{\partial^{2}¥\ta }{\partial T^{2}¥}=-\nabla_{\car}\left[
\mu_{0}({\bf R}-\ta(T))+U_{ext}(\car)-U_{ext}(\car-\ta(T))¥\right].
\label{eq:condkohn}
\ee
Since $\mu_{0}$ is position independent, it follows that the first 
term on the right hand side makes no contribution and we are left with 
\be
m\frac{\partial^{2}\ta}{\partial 
T^{2}¥}
=-\nabla_{\car}\left[ U_{ext}(\car)-U_{ext}(\car-\ta)\right].
\label{eq:kohn2}
\ee
For a harmonic trap potential described by
\be
U_{ext}(\car)=\frac{1}{2}m(\omega_{x}^{2}x^{2}+\omega_{y}^{2}y^{2}
+\omega_{z}^{2}z^{2}),
\ee
(\ref{eq:kohn2}) reduces to (\ref{eq:kohnosc}), as claimed.

We next argue that the rigid in-phase oscillation described by 
(\ref{eq:kohnde}) corresponds to the following distribution functions: 
\ba
&& f_{1}(\p,{\bf R},T)\equiv f_{10}(\p,{\bf R}-\ta(T))  
\nonumber \\
&& f_{2}(\p,{\bf R},T)\equiv f_{20}(\p,{\bf R}-\ta(T)),
\ea
where $f_{10}(\p,{\bf R})$ and $f_{20}(\p,{\bf R})$ satisfy the 
static equilibrium kinetic equations (\ref{eq:staticf1}) and 
(\ref{eq:f20}) and the static equilibrium densities satisfy 
(\ref{eq:kohnde}).
In this case, (\ref{eq:slowf1})  for $f_{1}({\bf p},{\bf R},T)$ 
reduces to
\ba
&&\left[ \frac{\partial}{\partial T}-\nabla_{\car}\left[ \nee(\car,T)+{\bf 
v}_{s}\cdot \p¥\right] \cdot \nabla_{\p}+\nabla_{\p}\left[ 
\nee(\car,T)+{\bf 
v}_{s}\cdot \p¥\right]\cdot \nabla_{\car}\right]f_{10}(\p,\car-\ta)
 \nonumber \\
&&=-ig\left[ m(\car,T)f_{20}(-\p,\car-\ta
)-m^{*}¥(\car,T)f_{20}(\p,\car-\ta) \right]+\nonumber \\
&&+\frac{g}{2}\left[\nabla_{\bf R}m({\bf R},T)\cdot \nabla_{\bf p}¥f_{2}(-{\bf p},{\bf R}-\ta)+
\nabla_{\bf R}m({\bf R},T)\cdot \nabla_{\bf p}¥f_{2}({\bf p},{\bf R}-\ta)\right].
\label{eq:f1kohn1}
\ea  
From the definition of $f_{i0}({\bf p},{\bf R})$, we see that
$f_{i0}(-\p,\car-\ta(T))=f_{i0}(\p,\car-\ta(T))$.
We also note that for the in-phase oscillation under consideration, 
$m(\car,T)=m_{0}(\car-\ta (T))$, where $m_{0}$ is real. 
Therefore the first term on the right side of (\ref{eq:f1kohn1}) vanishes, 
leaving us with
\ba
&&\left[ \frac{\partial}{\partial T}-\nabla_{\car}\left[\nee(\car,T)+{\bf 
v}_{s}\cdot \p \right] \cdot \nabla_{\p}+\nabla_{\p}\left[\nee(\car,T)+{\bf 
v}_{s}\cdot \p\right]\cdot \nabla_{\car}\right]f_{10}(\p, \car-\ta 
)\nonumber \\
&&=g\nabla_{\bf R}m_{0}({\bf R}-\ta)\cdot \nabla_{\bf p}f_{20}({\bf p},
{\bf R}-\ta).
\label{eq:f1kohn2}
\ea
Using (\ref{eq:energy}), (\ref{eq:cmpotential}) and (\ref{eq:kohnde}), 
it is straightforward to see that  $\nee({\bf R},T)=\nee({\bf R}-\ta) .
\label{eq:eneequiv}$
Using this, (\ref{eq:f1kohn2}) can be rewritten as
\ba
&&\left[ \frac{\partial}{\partial T}-\nabla_{\car}\nee(\car-\ta) 
\cdot \nabla_{\p}+\nabla_{\p}\left[\nee(\car-\ta)+\dot{\ta}\cdot {\bf 
p}\right] \cdot \nabla_{\car}\right]f_{10}(\p, \car-\ta 
)\nonumber \\
&&=g\nabla_{\bf R}m_{0}({\bf R}-\ta)\cdot \nabla_{\bf p}f_{20}({\bf 
p},{\bf R}-\ta).
\label{eq:f1kohn3}
\ea
If we introduce new variable ${\bf R'}(T)={\bf R}-\ta(T)$, and note 
that 
\be
\nabla_{\bf R}=\nabla_{\bf R'}\hspace{10mm} \frac{\partial}{\partial 
T}=-\dot{\ta}\cdot\nabla_{\bf R'},
\label{eq:newvar}
\ee
then (\ref{eq:f1kohn3}) can be 
rewritten as
\ba
\left[-\dot{\ta}\cdot \nabla_{\bf R'}-\nabla_{\bf R'}\nee({\bf R'}) \cdot 
\nabla_{\p}\right.&+&\left.\nabla_{\p}\nee({\bf R'}) 
\cdot \nabla_{\bf R'}+\dot{\ta}\cdot \nabla_{\bf R'}\right]f_{10}(\p,{\bf R'} ) 
\nonumber \\
&&=g\nabla_{\bf R'}m_{0}({\bf R'})\cdot \nabla_{\bf p}f_{20}
({\bf p},{\bf R'}).
\label{eq:f1kohn4}
\ea
Since the first and the last term on the left side of 
(\ref{eq:f1kohn4}) 
cancel each other, we are left with an equation which is precisely the 
same as the static HFB equation (\ref{eq:staticf1}) for 
the diagonal distribution function $f_{1}$.  

Similarly, for the in-phase mode, Eq.(\ref{eq:slowf2}) for the off-diagonal 
distribution function $f_{2}$ reduces to 
\be
\left[\frac{\partial}{\partial T}+2i\nee({\bf R},T)+ 
\dot{\ta} \cdot \nabla_{\bf R}\right] f_{20}(\p,\car-\ta) 
=-ig m_{0}(\car-\ta) \left[2f_{10}(\p,\car-\ta)+1\right].
\label{eq:f2kohn1}
\ee 
Using (\ref{eq:newvar}), (\ref{eq:f2kohn1}) becomes
\be
f_{20}(\p,{\bf R'}) 
=-\frac{g m_{0}({\bf R'})}{2\nee({\bf R'})}\left[2f_{10}(\p,{\bf R'})
+1\right].
\ee
Again, this result is seen to be equivalent to the static HFB equation result 
given by (\ref{eq:staticf2}) and (\ref{eq:f20}). 

In summary, we have explicitly verified that the ansatz given in 
(\ref{eq:kohnde}) satisfies our coupled HFB kinetic equations for the two 
distribution function $f_{1}$ and $f_{2}$. 
The oscillating center-of-mass displacement $\ta(T)$  satisfies the SHO equation of motion 
in (\ref{eq:kohnosc}). Thus, we have verified that our HFB coupled kinetic  equations 
exhibit a solution corresponding to a rigid SHO oscillation of the 
static equilibrium density profiles 
($n_{c},\tilde{n}$ and $\tilde{m}$) in the direction of 
$\eta_{\alpha}$, with the trap frequency $\omega_{\alpha}$. 
This expected solution is an important check on the correctness of our HFB equations of 
motion. We might note  that we did 
not make use of the approximate semi-classical local equilibrium forms 
for $f_{1}$ and $f_{2}$ given in Section IV to 
prove that the Kohn mode solution exists.  \\

\section{Conclusions}
To summarize, in this paper, we have used the HFB self-energy 
approximation to derive the equation of motion for the condensate 
order parameter (given by (17)) and the kinetic equations for the 
diagonal and the off-diagonal distribution functions for the 
non-condensate atoms (given by (\ref{eq:slowf1}) and (\ref{eq:slowf2})).
We have only solved these coupled equations to exhibit the rigid 
in-phase Kohn mode (Section V). However, we emphasize that, more generally, 
these equations will lead to collective modes which satisfy various 
conservation laws \cite{hm}, even though they were generated from the HFB 
single-particle self-energies (which  gives a single-particle 
spectrum with an energy gap in 
a uniform Bose gas). The essential physics is discussed in Sections 6 
and 7 of Griffin's Varenna lectures \cite{varenna}. Another way of 
saying this is that for a uniform Bose gas, the coupled equations of motion in this paper 
will lead to the same ``conserving'' density response function 
discussed by Cheung and Griffin \cite{cheung}.

Clearly the next step is to extend our present analysis by including the 
appropriate 
second-order self-energy contributions and hence, incorporate the effect of 
collisions into our kinetic equations. This generalization has been carried out for a uniform 
Bose-condensed gas by Kane and Kadanoff (KK) \cite{kk}, work which has 
been recently extended to trapped Bose gases \cite{milmsc}. KK concentrated on 
using their kinetic equations to derive the two-fluid hydrodynamics equations. 
Kirkpatrick and Dorfman \cite{kd} also derived 
a kinetic equation for the distribution function describing excitations
 in a uniform Bose-condensed gas,  using a 
different formalism  than KK. We believe that the KK approach 
based on non-equilibrium Green's functions gives the most systematic approach to 
deriving generalized quantum Boltzamnn equations for trapped Bose 
gases. 

\acknowledgements
We thank Tetsuro Nikuni  and Eugene Zaremba for useful comments on 
the manuscript.
This work was supported by a research grant from NSERC.  M.I.T. would 
also like to acknowledge a graduate scholarship from NSERC. 


\end{document}